\documentclass[twocolumn,superscriptaddress,amsmath,amssymb,aps,longbibliography,prx]{revtex4-2}
\usepackage[english,american]{babel}
\usepackage[colorlinks=true,urlcolor=blue,citecolor=blue,linkcolor=blue]{hyperref}
\usepackage{txfonts}
\usepackage{graphicx}
\usepackage{dsfont}
\usepackage{bbm}
\usepackage{booktabs}
\usepackage{amsmath}
\usepackage{algorithm}
\usepackage{algorithmicx}
\usepackage{algpseudocode} 
\usepackage{graphicx}% Include figure files
\usepackage{dcolumn}% Align table columns on decimal point
\usepackage{rotating}
\usepackage{longtable, booktabs, threeparttablex}
\usepackage{colortbl}

\usepackage{multirow}
\usepackage{threeparttable}
\usepackage{color}
\usepackage{xcolor} % Include the package
\definecolor{lightgreen}{RGB}{144,238,144} % RGB values for light green
\definecolor{lightgray}{gray}{0.75}

\newcommand{\Eq}[1]{Eq.~(\ref{#1})}

\begin{document}

%\title{Reinforcement learning \texttt{CrystalFormer} for more stable and property optimized materials}
%\title{Reinforcement learning from property prediction models for materials design}
%\title{Reinforcement Fine-Tuning for Stable and High-Performance Crystals}
\title{Reinforcement Fine-Tuning for Materials Design}

\author{Zhendong Cao}
\affiliation{Beijing National Laboratory for Condensed Matter Physics and Institute of Physics, \\Chinese Academy of Sciences, Beijing 100190, China}
\affiliation{School of Physical Sciences, University of Chinese Academy of Sciences, Beijing 100190, China}

\author{Lei Wang}
\email{wanglei@iphy.ac.cn}
\affiliation{Beijing National Laboratory for Condensed Matter Physics and Institute of Physics, \\Chinese Academy of Sciences, Beijing 100190, China}
\affiliation{Songshan Lake Materials Laboratory, Dongguan, Guangdong 523808, China}

\date{\today}

\begin{abstract}
Reinforcement fine-tuning played an instrumental role in enhancing the instruction-following and reasoning abilities of large language models. In this work, we employ reinforcement fine-tuning for materials design, in which  discriminative machine learning models are used to provide rewards to the autoregressive transformer-based materials generative model \texttt{CrystalFormer} [\href{https://www.sciencedirect.com/science/article/pii/S2095927325009752}{Sci. Bull. \textbf{70}, 3522 (2025)}]. By optimizing the reward signals—such as energy above the convex hull and material property figures of merit—reinforcement fine-tuning infuses knowledge from discriminative models into generative models. The resulting model, \texttt{CrystalFormer-RL}, shows enhanced stability in generated crystals and successfully discovers crystals with desirable yet conflicting material properties, such as substantial dielectric constant and band gap simultaneously. Notably, we observe that reinforcement fine-tuning not only enables the property-guided material design but also unlocks property-based material retrieval behavior of pretrained generative model. The present framework opens an exciting gateway to the synergies of the machine learning ecosystem for materials design.
\end{abstract}

\maketitle

\section{Introduction}

In materials science, machine learning (ML) techniques are revolutionizing research to enable rapid design and discovery of materials with desired properties. Machine learning models can be broadly categorized into generative and discriminative approaches. Crystal generative models, aim to capture the prior probability distribution $p(x)$ of stable crystal structures in the chemical space, learning the underlying patterns and correlations in crystal data to generate novel structures. These generative models~\cite{xie2021crystal, jiao2023crystal, zeni2023mattergen, gruver2024finetuned, cao2024space, jiao2024space} capture the underlying distribution of stable materials in the chemical space and enable the direct generation of novel materials. On the other hand, discriminative models, including machine learning interatomic potentials (MLIP)~\cite{unke2021machine, PhysRevLett.120.143001, batzner20223, batatia2022mace, zhang2022dpa, zhang2023dpa2, neumann2024orb} and property prediction models~\cite{PhysRevLett.120.145301, chen2019graph}, focus on modeling the conditional probability $p(y|x)$, where $y$ represents material properties given a crystal structure $x$. These models excel at predicting specific properties or behaviors of materials and enable accelerated simulations and property predictions. The integration of these complementary approaches—combining the generative capability of models that capture $p(x)$ with the predictive power of discriminative models that capture $p(y|x)$—presents a powerful framework for materials discovery. Further achievements hinge on the flexible combination of discriminative and generative models, enhancing their performance and adapting them to specific application domains. Reinforcement learning (RL) provides a principled way to achieve this synergy, allowing the generative model to be improved and guided by the knowledge embedded in discriminative models.

%\lw{I feel start the discussion with supervise/unsupervise learning is a bit too general. 
%Maybe start from crystal generative models (diffusion, autoregressive), then move on to the need of fine tune for either better poformance or domain specification. Move some of the supervise/unsupervised learning to the discussion section. }
% Machine learning (ML) is revolutionizing the field of materials science, enabling the rapid discovery of novel materials with desired properties.
% There are two principal methodologies in the field of machine learning in materials science: supervised and unsupervised learning. The former entails the training of models on labeled data, such as property prediction models~\cite{PhysRevLett.120.145301, chen2019graph} and machine learning interatomic potential (MLIP)~\cite{unke2021machine}, to predict specific material properties.
% In contrast, unsupervised learning methods, such as crystal generative models~\cite{xie2021crystal, jiao2023crystal, jiao2024space, zeni2023mattergen, gruver2024finetuned, cao2024space}, aim to learn the underlying distribution of materials in the chemical space, enabling the generation of novel materials with desired properties.
% The rapid development of both supervised and unsupervised learning methodologies has significantly advanced the field of materials informatics. However, the integration of these two methodologies remains a critical challenge.

The synergy of discriminative and generative machine learning models has already shown great success in real-world applications in the past few years. 
A notable example is classifier-guided generation~\cite{dathathri2019plug, dhariwal2021diffusion}, where supervised trained discriminative models are leveraged to guide sampling of generative models, eliminating the need for repeated retraining of the generative model. Moreover, in the post-training phase of large language models (LLM), reinforcement learning from human feedback (RLHF)~\cite{ouyang2022training, ziegler2019fine} is employed to extract signals from supervised trained reward
models to instruct generative models. In this context, RL~\cite{sutton2018reinforcement} serves as a powerful mechanism to integrate unsupervised and supervised learned models. For example, LLM trained via RL have demonstrated improved instruction following~\cite{ouyang2022training} and reasoning capabilities~\cite{deepseekai2025deepseekr1}, which shows it is an effective way to enhance the performance of generative models.

Inspired by RLHF, we devise an RL approach to finetune the crystal generative model \texttt{CrystalFormer} using the reward provided by discriminative models such as MLIP and property prediction models, see Fig.~\ref{fig:workflow}(a).
We demonstrate that the approach effectively enhances the stability and desired figures of merit of generated materials. We have released the codes and trained models~\cite{github}. Moreover, we provide instructions on fine-tuning \texttt{CrystalFormer} with customized MLIP or property prediction models for broader applications. 

%This integrated method aims to accelerate the discovery of novel materials with desired properties, thereby advancing the field of materials science. 
%\lw{I commented out the previous sentence because it is an empty statement. Let us avoid them in the whole paper. You may say it once of twice (in the abstract and ending paragraph), but not more.}
%\zdcao{OK!}

The organization of the paper is as follows: Section~\ref{sec:method} describes the RL algorithm for fine-tuning \texttt{CrystalFormer}. Section~\ref{sec:results} presents the results on improving the stability of generated materials and examples of property-controlled material generation. Section~\ref{sec:discussion} discusses the prospects of this work in a broader context.

\section{Method}
\label{sec:method}

\begin{figure}[t!]
    \centering
    \includegraphics[width=1\linewidth]{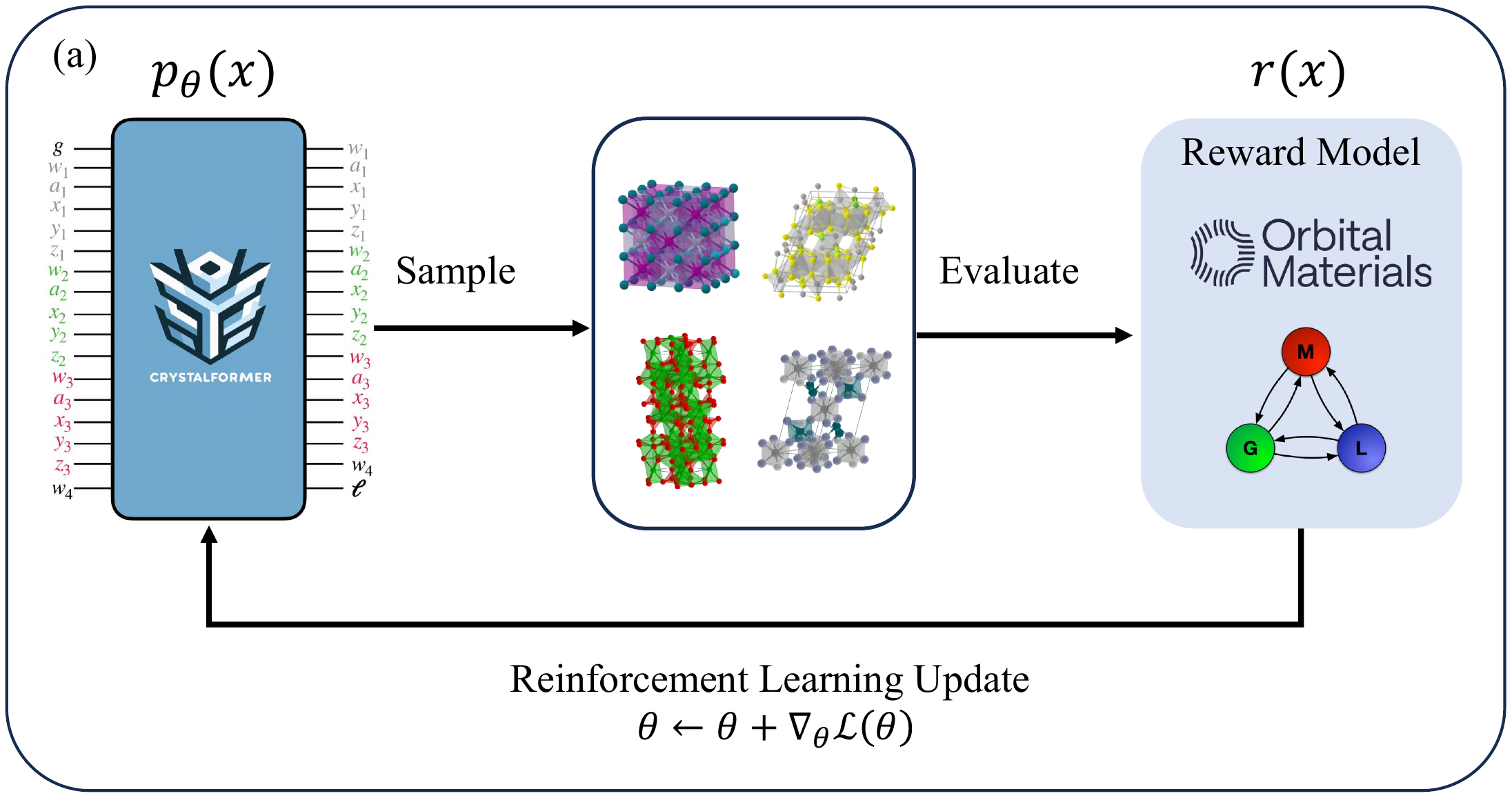}
    \includegraphics[width=1\linewidth]{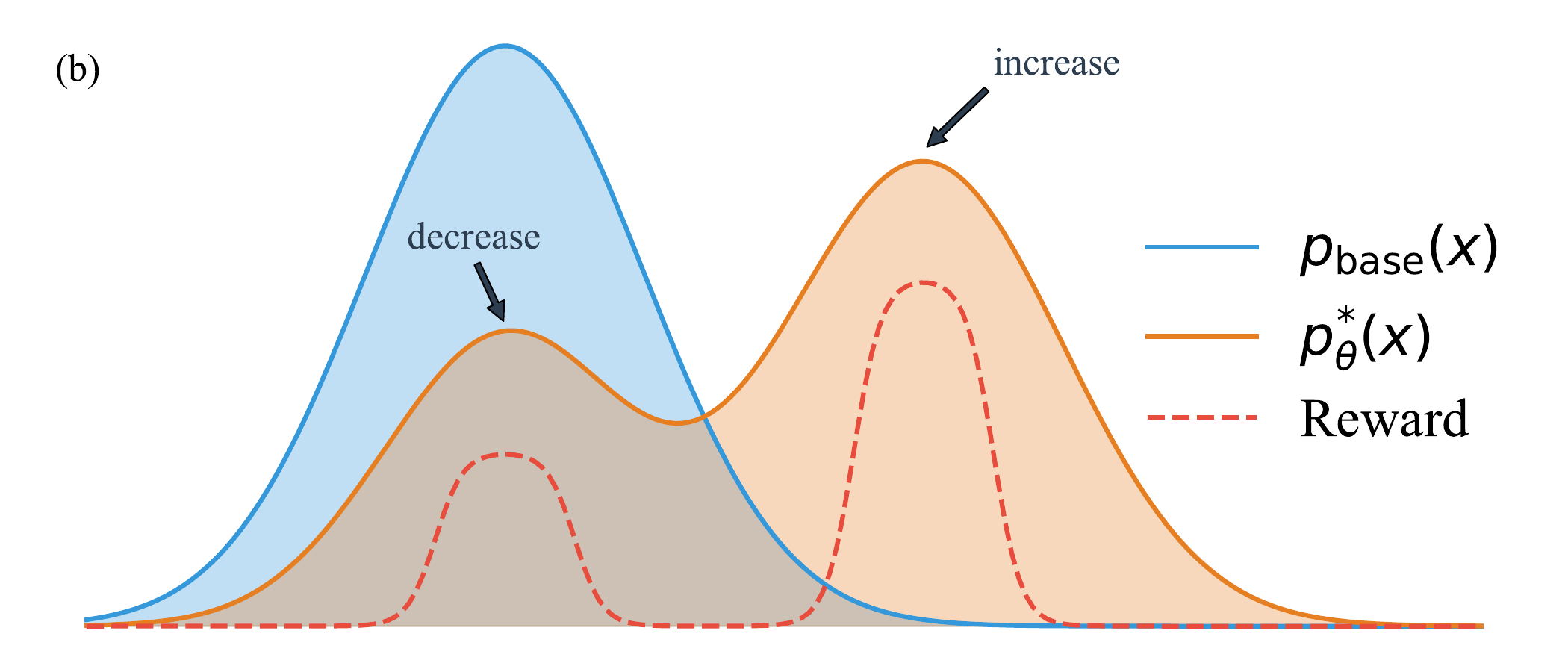}
    \caption{(a) The reinforcement fine-tuning workflow. Machine learning interatomic potential or property prediction models 
    provide rewards to the material generated by \texttt{CrystalFormer}. The RL training loop updates the parameters of a pre-trained \texttt{CrystalFormer} to maximize the objective function in \Eq{eq:rlhf}. 
    (b) The reinforcement fine-tuned model deviates from the base model to maximize the expected reward with entropy regularization. 
    }
    \label{fig:workflow}
\end{figure}

In the spirit of RLHF~\cite{ziegler2019fine, ouyang2022training}, we fine tune pre-trained crystal generative model using reward signal from MLIP or a material property prediction model. The objective function to be maximized reads~\cite{schulman2017equivalence},
\begin{equation}
    \mathcal {L} = \mathop{\mathbb{E}}_{x \sim p_{\theta}(x)} \left [r(x) - \tau \ln \left( \frac{p_{\theta}(x)}{p_{ \mathrm{base} }(x)} \right) \right ],
\label{eq:rlhf}
\end{equation}
where $x$ represents crystalline materials sampled from a policy network $p_{\theta} (x)$, $r(x)$ is the reward function which awards preferred materials with high returns, $\tau$ is the regularization coefficient controls proximity to the base model $p_{\mathrm{base}}(x)$. The second term of \Eq{eq:rlhf} is the Kullback-Leibler (KL) divergence between the policy distribution and the base model~\cite{jaques2017sequence}. Overall, the objective function aims to maximize the expected reward of samples while ensuring that the policy also aligns with the base model. 

The combination of both terms in \Eq{eq:rlhf} can also be regarded as a KL divergence between the policy network $p_\theta(x)$ under fine-tuning and the optimal policy~\cite{schulman2017equivalence, korbak2022rl}
\begin{equation}
    p^{*}_{\theta}(x) = \frac{p_{\mathrm{base}}(x)}{Z}  \exp \left (\frac{r(x)}{\tau}  \right ),
\label{eq:optimal_policy}
\end{equation}
where $Z$ is the normalization factor. The corresponding optimal value of the objective function \Eq{eq:rlhf} will be $\tau \ln Z$. Setting the reward to be proportional to the log-likelihood of certain desired property $r(x) \propto \tau \ln p(y|x)$, one sees that $p^{*}_{\theta}(x)  \propto p_{\mathrm{base}}(x)  p(y|x) $. Therefore, \Eq{eq:optimal_policy} can be viewed as Bayesian inference with a prior distribution $p_{\mathrm{base}}(x)$ and the likelihood $\exp \left (r(x)/ {\tau} \right )$~\cite{levine2018reinforcement, korbak2022rl}. In practice, the pre-trained model not only provides the prior distribution $p_{\mathrm{base}}(x)$ but also provides parameter initialization for the policy network. Figure~\ref{fig:workflow}(b) illustrates the RL fine-tuning process. The variational inference provides an alternative to the Markov chain Monte Carlo (MCMC) sampling carried out in Ref.~\cite{cao2024space}. Here, larger $\tau$  corresponds to higher temperature and hence stronger entropy regularization. Such optimization is also equivalent to the variational free energy calculations carried out for statistical mechanics problems~\cite{neuralrg, van}, where the two terms in the objective function \Eq{eq:rlhf} correspond to expected energy and entropy, respectively. We employ the proximal policy optimization (PPO)~\cite{schulman2017proximal} algorithm to maximize the expected reward. The implementation details are provided in Appendix~\ref{sec:ppo-app}. 

In the RLHF paradigm for LLM, a reward model that represents human preferences is used to evaluate the quality of generated samples. Analogously, in computational material science, density functional theory (DFT) is
frequently employed to assess the quality of crystalline materials. However, the computational expense associated with DFT renders its direct application as a reward model within the RLHF framework infeasible.
To circumvent this limitation, we propose the utilization of an MLIP as a surrogate reward model. Recent advancements in MLIP have yielded models capable of simulating
atomic systems with both high accuracy and reduced computational cost, thus providing a viable proxy for DFT calculations~\cite{unke2021machine}. Moreover, universal MLIP~\cite{chen2022universal, batatia2024foundation, neumann2024orb, zhang2022dpa, zhang2023dpa2, yang2024mattersim}, encompassing the entire periodic table of chemical elements, offering
the capacity to evaluate a diverse range of materials, making them particularly well-suited for our RL fine-tuning. %Consequently, employing a universal MLIP as the reward model obviates the need for iterative reward model retraining, as is typically required in RLHF for large language models.

RLHF can also be effectively employed for property-guided material generation, wherein the reward model is one or a combination of several property prediction models~\cite{park2024inverse, thomas2025reinforceing, Popova_2018, Mazuz2023}.
The fine-tuned generative model is capable of generating materials with desired properties, such as band gap, formation energy, and so on. We choose to carry out RL fine-tuning for the crystal generative model \texttt{CrystalFormer}~\cite{cao2024space} due to its simplicity, generality, and flexibility. 

The \texttt{CrystalFormer} is an autoregressive transformer which models the tokenized representation of crystal structures with explicit knowledge of space group symmetries. The representation includes the space group number, Wyckoff letter, chemical element, and fractional coordinates of each symmetry inequivalent atom, and finally, the lattice parameters~\cite{cao2024space}. For example, the crystal LaH$_{10}$ in the $Fm\bar{3}m$ (No.~225) space group with the cubic conventional unit cell of the length $5.1\AA$ is represented as the sequence "{225-a-La-0-0-0-c-H-1/4-1/4-1/4-f-H-0.375-0.375-0.375-5.1-5.1-5.1-90-90-90}", which contains all the necessary information of the crystal. Key to the design of \texttt{CrystalFormer} is the sequential nature of Wyckoff letters which decrease in site symmetry in an alphabet order. In this regard, the \texttt{CrystalFormer} leverages Nature's codebook--the Wyckoff position table--for a multimodal probabilistic modeling. Although the representation appears to lack a sense of spatial geometry, unsupervised pre-training allows the model to ingest solid-state chemistry knowledge by compressing the crystalline material database into the model parameters. Furthermore, the RL phase may allow it to extract the geometry and physics from MLIP and property prediction models trained with supervised approaches.

In the present work, we first train the \texttt{CrystalFormer} on the Alex-20 dataset, which is curated from the Alexandria dataset~\cite{schmidt2024improving}. More details of the dataset can be found in the Appendix~\ref{sec:dataset-app}.
The resulting pre-trained model then serves as the base model for the RL procedure, where we fine-tune the model using the RL method using the MLIP or crystal property prediction models as the reward model. 
In the RL stage, the model learns from the reward model on its own samples. We sample crystal structures from \texttt{CrystalFormer}'s policy, $p_{\theta}(x)$, with the space groups distributed according to the Alex-20 dataset. The samples are then evaluated by the reward model $r(x)$, which assigns a reward based on stableness or properties. These reward signals provide feedback for updating the \texttt{CrystalFormer}'s policy to maximize the objective function in \Eq{eq:rlhf}, iteratively improving the \texttt{CrystalFormer}'s capacity to generate target crystal structures. We stop the fine-tuning process when the loss function converges.

\section{Applications}
\label{sec:results}

\subsection{Reinforcement learning from MLIP for improved stability of generated materials}
\label{sec:stability}

\begin{figure}[t!]
    \centering
    \includegraphics[width=1\linewidth]{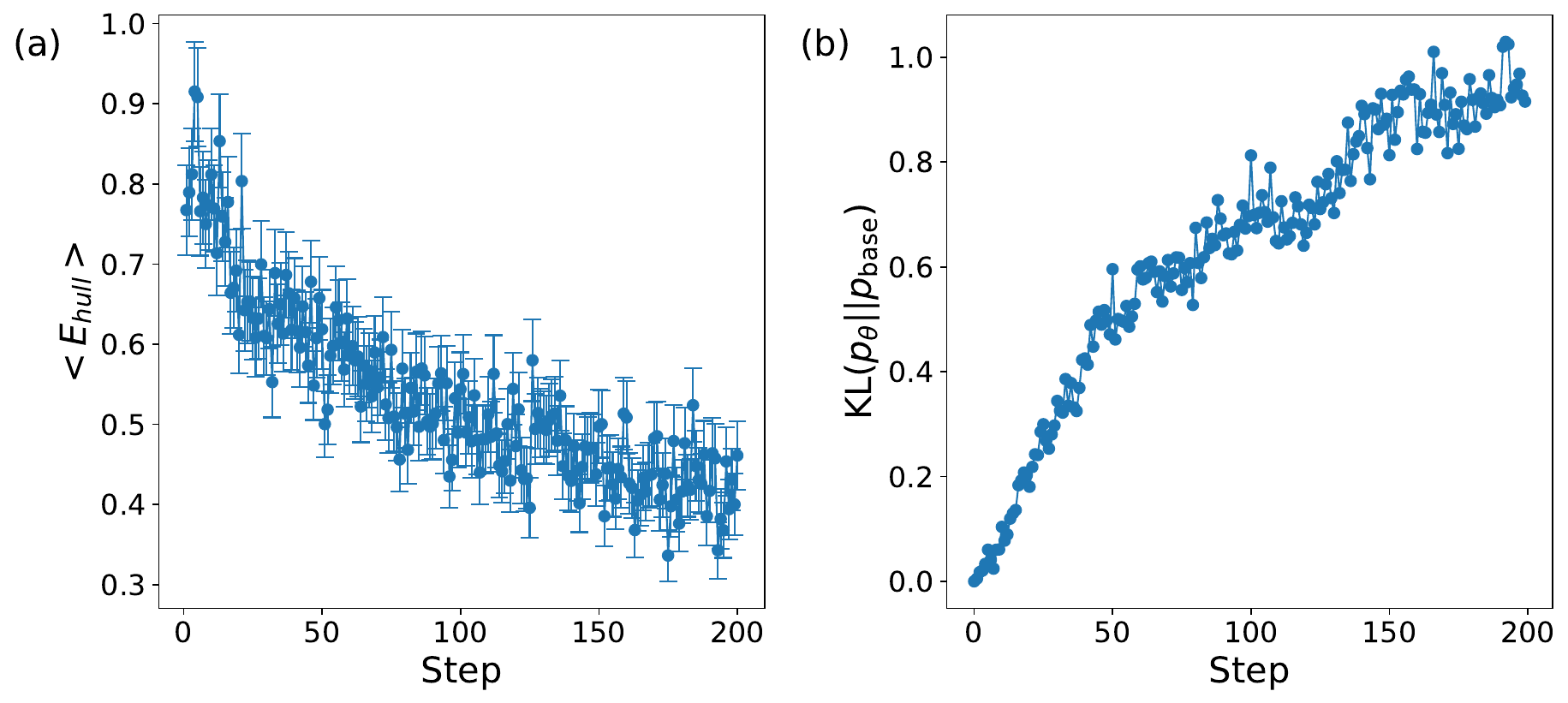}
    \caption{(a) The average energy above the convex hull of generated materials
    and (b) the KL divergence between the policy and the base model versus training steps. We set the regularization coefficient $\tau=0.1$.
     }
    \label{fig:reward}
\end{figure}

As the first application of the reinforcement fine-tuning of \texttt{CrystalFormer}, we choose the energy above the convex hull as the reward signal, i.e., $r(x) = -E_{\mathrm{hull}}(x)$. The energy above the convex hull is the key metric for assessing the stability of a material, with lower values indicating greater stability.
To predict the energy and calculate the energy above the convex hull based on the Alexandria convex hull~\cite{schmidt2024improving}, we utilize 
the Orb model~\cite{neumann2024orb}. This choice allows us to obtain accurate energy predictions and assess the stability of materials relative to the convex hull fast.
It is important to emphasize that our approach is not limited to the Orb model; any universal MLIP that can compute energy can serve as the reward model. More accurate and robust MLIP yields better reward signals, enhancing alignment with DFT calculations. Benchmarks on universal MLIP such as~\cite{riebesell2024matbench} provide an interactive leaderboard to select the favorable one for the fine-tuning purpose. We evaluate the performance by measuring the fraction of structures that are stable, unique, and novel (S.U.N.)~\cite{zeni2023mattergen}.
The stability is defined as the energy above the convex hull being less than 0.1 eV/atom, uniqueness is defined as the generated structure not being identical to any other generated structures, and novelty is defined as the generated structure not being present in the training dataset.

Figure~\ref{fig:reward}(a) shows the expected energy above the convex hull of generated samples, which shows steady improvement during the training process. Figure~\ref{fig:reward}(b) shows the KL divergence between the policy and the pre-trained base model $\mathrm{KL}(p_\theta \Vert p_\mathrm{base}) = \mathop{\mathbb{E}}_{x \sim p_{\theta}(x)} \left [  \ln \left( \frac{p_{\theta}(x)}{p_{ \mathrm{base} }(x)} \right )\right ] $, which is orders of magnitude smaller than change in the negative log-likelihood during pre-training~\cite{cao2024space}.
%\lw{or maybe something else that are more informative for the training dynmaics ?  number of elements ? number of atoms in the unit cell ? Will there be interesting findings ?  I mention this because  the lenght of response keep increasing in deepseek R1 zero during RL.}
%\zdcao{the increasing of response length always happens in DPO training, but PPO is more stable.}

\begin{figure}[t!]
    \centering
    \includegraphics[width=1\linewidth]{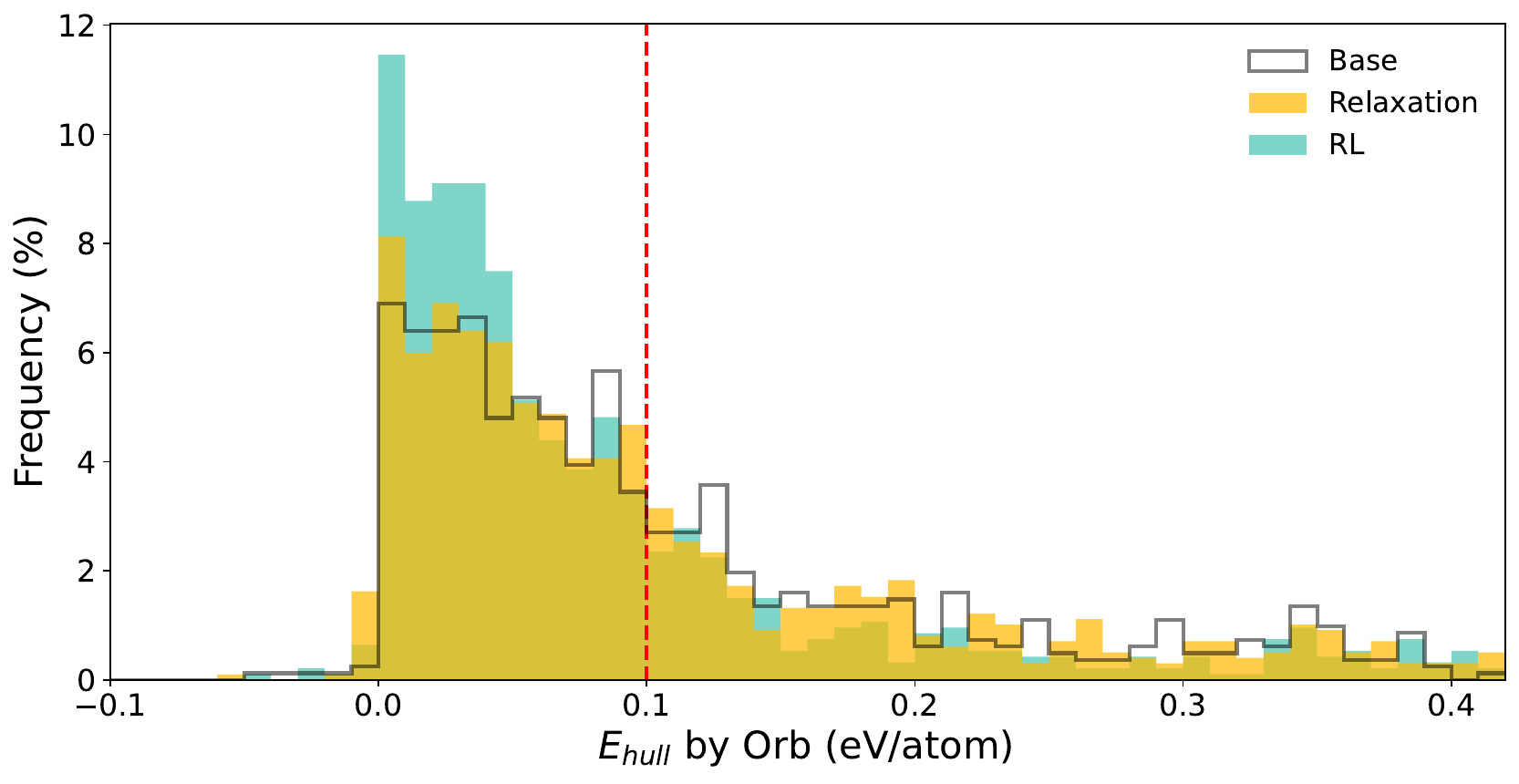}
    \caption{The histogram of energy above convex hull for the crystal samples from the pre-trained base model and the ones relaxed
    by the Orb model~\cite{neumann2024orb}. 
    In comparison, reinforcement fine-tuned of the model significantly reduces the energy above convex hull of generated materials.
    The red dashed line indicates the threshold of 0.1 eV/atom for stable materials. 
    Relaxation ratio from 44.7\% in the base model to 57.8\%, while RL fine-tuning improves the ratio of stable materials to 73.4\% even without relaxation.
    }
    \label{fig:ehull}
\end{figure}

\begin{figure*}[t!]
    \centering
    \includegraphics[width=\linewidth]{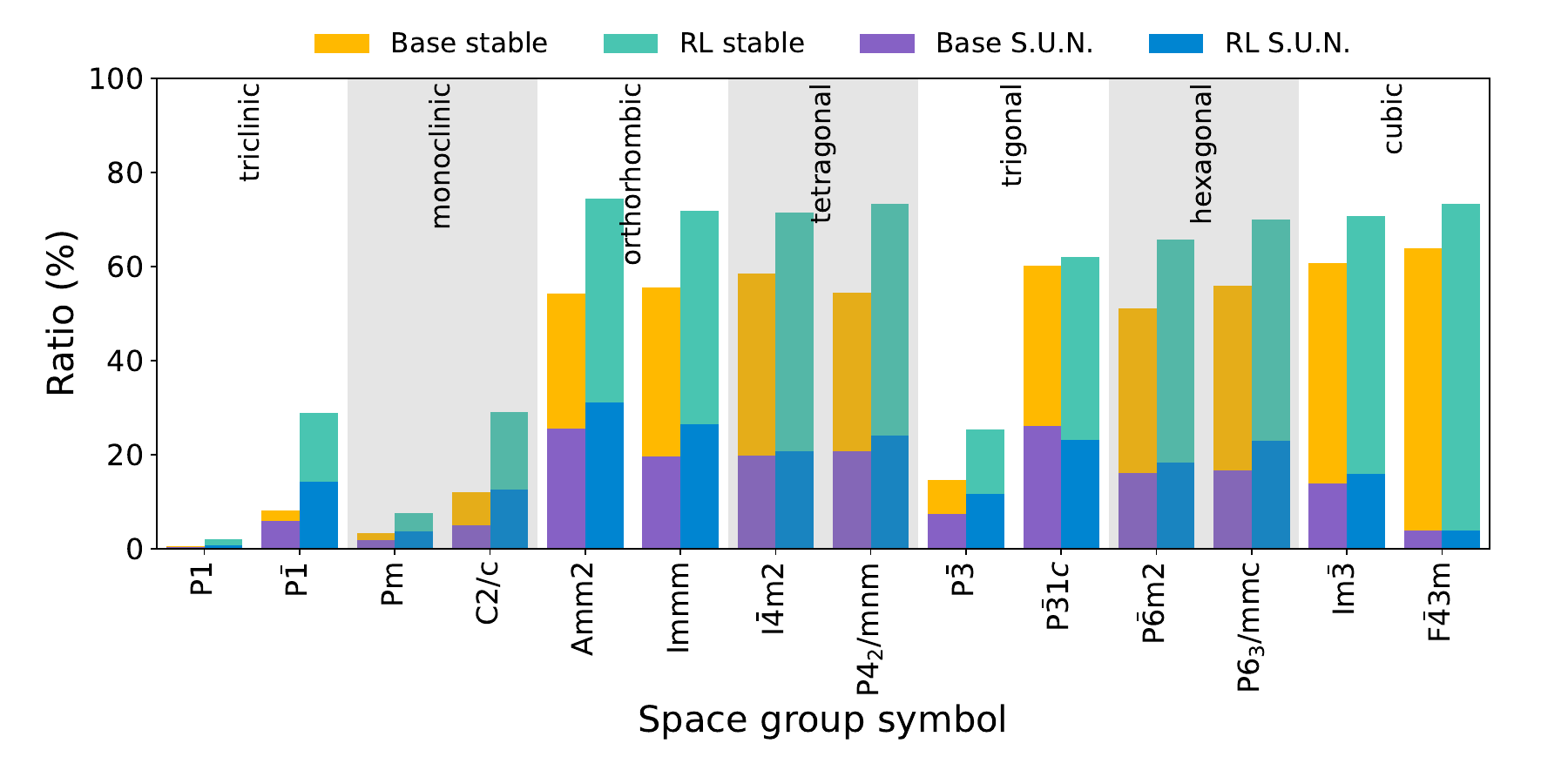}
    \caption{Fraction of generated structures that are stable, novel and unique (S.U.N.) for 14 space groups spanning the seven crystal systems.
    For each space group, two side-by-side bars correspond to the samples from the pre-trained (``Base'') and RL fine-tuned (``RL'') models.
    The S.U.N. structures form a subset of stable materials each space group. The shaded regions separate different crystal systems.
    }
    \label{fig:sun_ratio}
\end{figure*}

Figure~\ref{fig:ehull} shows the histogram of the energy above the convex hull for the pre-trained model and the reinforcement fine-tuned model. 
For each model, 1000 structures were sampled, with their space groups drawn from the training data distribution.
% We note that we do not relax the generated structures, as the focus of the fine-tuning is to enhance the stability of generated materials.
For the pre-trained model, 44.7\% of the generated structures show $E_{\mathrm{hull}}<0.1$ eV/atom. Relaxation improves this ratio to 57.8\%. 
On the other hand, reinforcement fine-tuning with respect to the energy above the convex hull enhances the ratio to 73.4\%. Such substantial improvement is due to that the reinforcement fine-tuning reshapes the distribution with a global modification of the
Wyckoﬀ sites, fractional coordinates, and lattice parameters while relaxation only modifies the materials locally. Besides reducing the energy above the convex hull, the reinforcement fine-tuning also improves the structure validity metrics compared to the base model~\cite{xie2021crystal}, see Appendix~\ref{sec:dataset-app}.

Figure~\ref{fig:sun_ratio} shows the fraction of generated structures that are stable and are also S.U.N. for 14 space groups spanning the seven crystal systems.
We investigate the performance of the model on the same 14 space groups as Ref.~\cite{zeni2023mattergen}. For each space group, we generate 1000 structures and evaluate the quality of samples using the Orb model~\cite{neumann2024orb}. 
The results demonstrate that the reinforcement fine-tuned model generates a higher fraction of stable and S.U.N. structures in most cases, indicating that the RL method can enhance the model's ability to generate S.U.N. structures even though we only award for the stability.
% For space groups with high symmetries (e.g. $F\bar{4}3m$), even though the fraction of stable materials increases, the S.U.N. ratio does not increase due to the restricted degrees of freedom limiting the structural motifs.
Reinforcement fine-tuning with respect to the energy above the convex improves the stability across all space group (as it should). However, the improvement of the S.U.N. is impaired by the novelty criteria as there is strong constraints to the structural motifs in these high-symmetry systems.
From the figure one also observes that the performance of \texttt{CrystalFormer} is worst for the least symmetric $P1$ space group where there is minimal symmetry-related inductive bias to be explored. This is in sharp contrast to Ref.~\cite{zeni2023mattergen} which shows the best S.U.N. in the $P1$ space group, highlighting the distinctive advantage of our method. In fact, for space group with higher symmetry which are more relevant to the natural distribution of inorganic materials~\cite{urusov2009frequency}, we observed that the performance of \texttt{CrystalFormer} is on par or even better than the ones shown in Fig. D8 of Ref.~\cite{zeni2023mattergen}. Lastly, while reinforcement fine-tuning does not affect the generation speed of the model, it increases the rate of generating S.U.N. samples from 3.46 to 4.89 per second. %Notably, we define the S.U.N. speed as the number of S.U.N. structures generated per second, which is a critical metric for practical applications.

%This issue can be alleviated by incorporating a diffusion or flow model~\cite{jiao2024spacegroup, sriram2024flowllm} after the \texttt{CrystalFormer} to refine the generated structures.

\begin{figure}[t!]
    \centering
    \includegraphics[width=\linewidth]{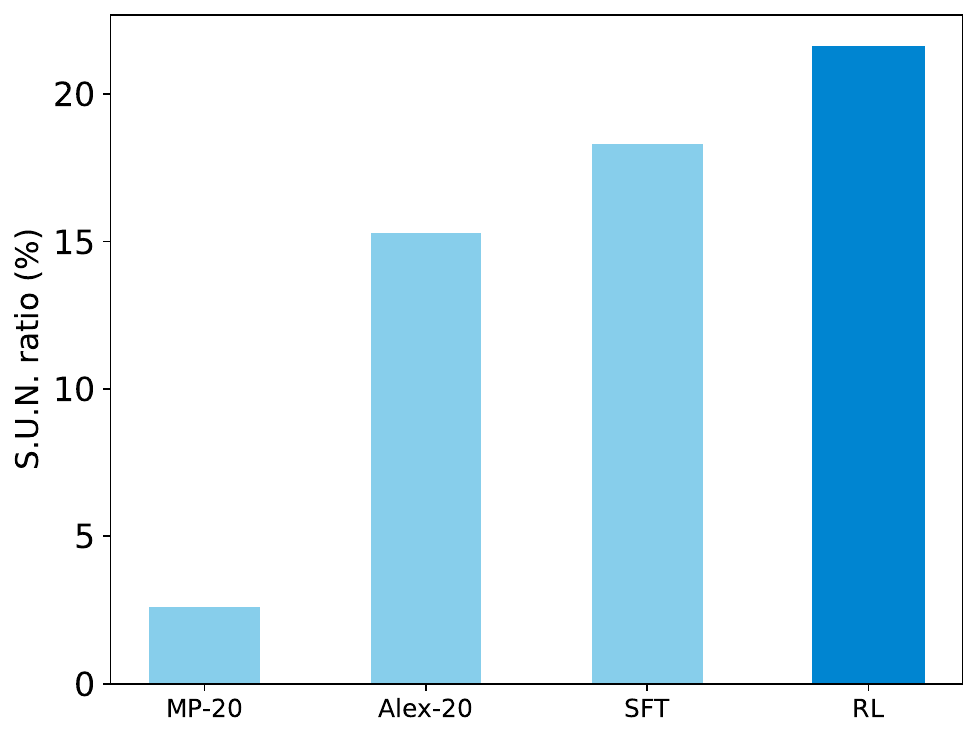}
    \caption{The weighted S.U.N. ratio averaged over all space groups of the \texttt{CrystalFormer} trained on the MP-20 dataset and the Alex-20 dataset compared with the fine-tuned model with SFT and RL approaches.
    The energy above the convex hull is calculated based on the Alexandria convex hull and the novelty is calculated based on the Alex-20 dataset.
    }
    \label{fig:rl_vs_pretrain}
\end{figure}

Figure~\ref{fig:rl_vs_pretrain} breaks down the performance enhancement in terms of the S.U.N. ratio coming from enlarging the pre-training dataset and different ways of fine-tuning. For these evaluations, we sample space group following the same distribution of the training dataset and use it as precondition to the \texttt{CrystalFormer}. First of all, one sees that enlarging the pre-training dataset from MP-20 to Alex-20 yields a significant improvement in the S.U.N. ratio, rising from 2.6\% to 15.3\%.
Next, we also tested the improvement brought by supervised fine-tuning (SFT) on the pre-trained model. For SFT, we generate 10,000 crystal samples and relax them using the Orb models. Only those stable samples with $E_{\mathrm{hull}}<0.1$ eV/atom were used to finetune the pre-trained model.
Figure~\ref{fig:rl_vs_pretrain} shows that SFT yields some improvement over the pre-trained model, further increasing the S.U.N. ratio from 15.3\% to 18.3\%.
However, it is the reinforcement fine-tuned model that delivers the most significant increase in the S.U.N. ratio, leading to a final S.U.N. ratio of 21.6\%.

%S.U.N. ratio across all generated structures increased from 15.3\% for the pre-trained model to 21.6\%  for the RL fine-tuned model.

\subsection{Reinforcement learning from the figure of merits for property-guided material generation}
\label{sec:property-guided}

\begin{figure}[htbp]
    \centering
    \includegraphics[width=\linewidth]{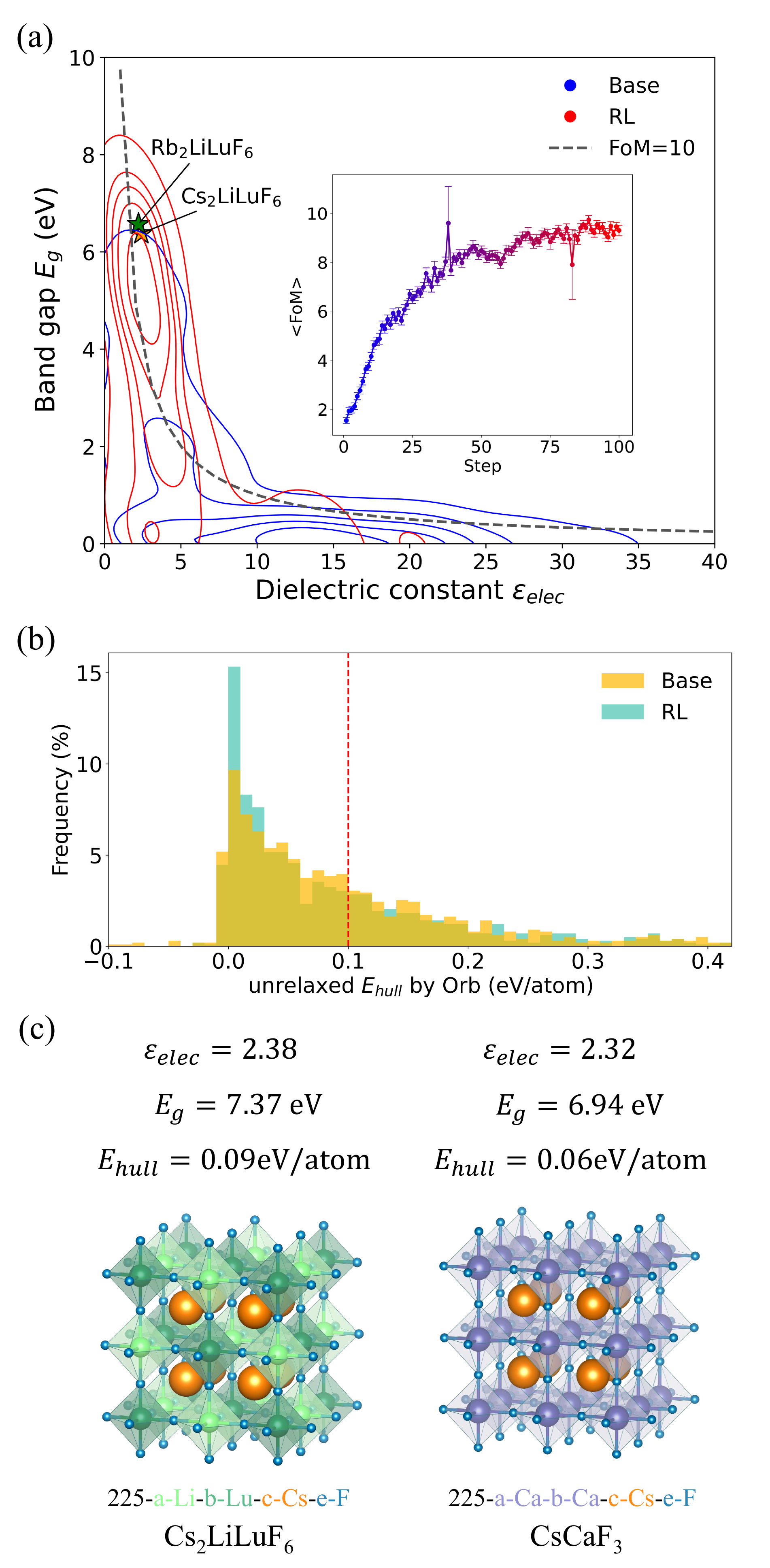}
    \caption{(a) The distribution of band gap $E_{g}$ and electronic dielectric constant $\varepsilon_{\mathrm{elec}}$ of material samples generated in the $Fm\bar{3}m$ space group (No.225). The two properties are predicted by the MEGNet and AnisoNet models, respectively.
    The dashed line indicates a constant FoM at values $ \varepsilon_{\mathrm{elec}} \cdot E_{g} =10 $eV. 
    % Note that we subsample 1000 data points from the generated samples for better visualization. 
    The inset shows  average FoM of generated materials versus training steps. Two stars indicate the S.U.N. materials.
    (b)  The histogram of energy above the convex hull predicted by the Orb model~\cite{neumann2024orb} for the crystal samples from the pre-trained and the RL fine-tuned model in $Fm\bar{3}m$ space group (No.225), respectively.
         The red dashed line indicates the threshold of 0.1 eV/atom and the $x$-axis is up to 0.4 eV/atom.
         RL improves the FoM of the generated samples towards high FoM dielectric materials at the cost of only slightly affecting the stability.
         (c) Two examples of discovered high FoM materials along with their properties verified by DFT calculation. We also show the tokenized material string generated with \texttt{CrystalFormer} (fractional coordinate and lattice parameters omitted). 
    } 
    \label{fig:dielectric}
\end{figure}

Next, we apply the RL method for dielectric material discovery.
% Dielectric materials are essential components in a wide range of contemporary electronic devices, including flash storage, central processing units, and random access memory~\cite{wang2018highk, doi:10.1021/cr9001275}.
% The functional utility of these materials is fundamentally governed by the precise interplay between the dielectric constant and band gap—two inherently anti-correlated properties that seldom achieve optimality concurrently in a single material.
% Consequently, delineating a Pareto front becomes essential for navigating the trade-offs between these competing parameters.
The interplay between the electronic dielectric constant and the band gap reflects a fundamental constraint in condensed matter systems.
Recent works have established that the electronic polarizability originates from virtual interband transitions governed by the quantum metric and inversely scales with the spectral gap~\cite{Komissarov2024}, while the gap itself is bounded by the dielectric response and electron density~\cite{PhysRevB.110.155107}.
Pursuing materials that simultaneously exhibit a large electronic dielectric constant and a wide band gap thus represents a key challenge in realizing highly polarizable yet robust insulators, where quantum geometry and energetic stability reach a delicate balance.
We utilize the pre-trained MEGNet~\cite{chen2019graph,chen2021learning} and AnisoNet~\cite{D4FD00096J} models to predict the band gap $E_{g}$ and electronic dielectric constant $\varepsilon_{\mathrm{elec}}$ of the generated samples, respectively.
MEGNet achieves a mean absolute error (MAE) of 0.314 eV for band gap prediction on the Materials Project dataset~\cite{jain2013commentary}, while AnisoNet achieves a MAE of 0.297 for electronic dielectric constant prediction on the Materials Project dielectric dataset~\cite{jain2013commentary, petousis2017high}.
It is worth noting both models are trained on the data computed using the Perdew-Burke-Ernzerhof functional~\cite{Perdew.PBE.1996}, which is known to underestimate the band gap.
These data also omit corrections such as spin-orbit coupling or excitonic effects, which may be important for certain materials.
More accurate band gap prediction model would require training on higher-fidelity datasets, such as those computed using hybrid functionals~\cite{heyd2003hybrid, heyd2006hybrid}, the GW method~\cite{PhysRevLett.96.226402} or GW-BSE approach~\cite{RevModPhys.74.601}.
The electronic dielectric constant is a dimensionless number, which is the ratio of the permittivity of a substance to the permittivity of vacuum.
% To predict the total dielectric constant $\varepsilon_{\mathrm{total}}$ which includes the ionic and electronic contributions~\cite{riebesell2024pushing}, we train a separate M3GNet~\cite{chen2022universal} model on the dataset of density functional perturbation theory (DFPT) calculations in the Materials Projects~\cite{jain2013commentary, petousis2017high},
% more details of training and evaluation can be found in Appendix~\ref{sec:dielectric-app}.
The reward function is defined as the figure of merit (FoM), $r(x) = \varepsilon_{\mathrm{elec}} \cdot E_{g}$, which favors materials with both sizeable band gap and permittivity. 

Figure~\ref{fig:dielectric}(a) shows the ML predicted band gap $E_{g}$ versus electronic dielectric constant $\varepsilon_{\mathrm{elec}}$ for generated samples in the $Fm\bar{3}m$ space group (No.225) using the base and fine-tuned model respectively. 
The reinforcement fine-tuning significantly improves the FoM, with the majority of generated samples exhibiting higher FoM values compared to the pre-trained base model.
We observe that the RL method primarily enhances the FoM by increasing the band gap, rather than the dielectric constant.
This outcome may be attributed to the limited and relatively low dielectric constant data used to train the dielectric model, potentially causing the model to predict lower dielectric constants.
To address this issue, a more robust dielectric prediction model that captures a wider range of dielectric constants could be employed~\cite{falletta2024unified}.
In contrast, the band gap prediction model demonstrates greater accuracy, resulting in the RL method favoring an increase in the band gap to optimize the FoM.

We first generated 1000 samples and then use Orb model to filter out the unstable materials, followed by DFT verification of the top 50 candidates with the highest FoM in the remaining subset.
More computational details are provided in Appendix~\ref{sec:computational-details}, and materials with $E_\mathrm{hull}<0.1$ eV/atom are listed in Table~\ref{table:dielectric_samples}.
We observed that many of these samples are fluoride, probably due to fluorine's strong electronegativity increases the ionic character of bonds, so that to ensure a sizeable band gap. Moreover, many of these high FoM samples are composed of Cs elements, which may be due to the relatively large ionic radius of Cesium, leading to higher polarizability and thus an increased total dielectric constant of the material.
Figure~\ref{fig:dielectric}(b) shows the energy above the convex hull for the materials sampled from the base and fine-tuned model respectively, which suggests that the reinforcement fine-tuning does not significantly affect the stability of generated materials.
Two examples of high FoM materials discovered by \texttt{CrystalFormer-RL}, Cs$_2$LiLuF$_6$ and CsCaF$_3$, are shown in Fig~\ref{fig:dielectric}(c).
Cs$_2$LiLuF$_6$ belongs to the double perovskite~\cite{wolf2021doubling} structure prototype.
% We note that although it is flagged as novel in Table~\ref{table:dielectric_samples}, it is related to partial substitution~\cite{merchant2023scaling} of the Ba element in CsBaF$_3$ which is in the Alex-20 dataset.
CsCaF$_3$ is obtained by substituting Ba with Ca in CsBaF$_3$, which is included in the Alex-20 dataset, and it crystallizes in the halide perovskite~\cite{wolf2021doubling, cai2019high} structure prototype.
% Therefore, the synthesis and identification of Cs$_2$BaSrF$_6$ as a novel compound may subject to caveats pointed out in Ref.~\cite{PRXEnergy.3.011002} such as compositional disorder.
% % The most similar structure to Cs$_2$LiBaF$_6$ found in the DFPT dataset is Cs$_2$LiInF$_6$, which exhibits a band gap of 5.23 eV and a total dielectric constant of 25.42, resulting in a FoM value of 132.95 eV.
% Li$_6$PbCl$_8$ is derived from Li$_6$VCl$_8$~\cite{HANEBALI1981887} by substituting V with Pb and belongs to the Suzuki structure prototype~\cite{suzuki1961x}.

Many of the generated samples are present in the Alex-20 dataset which is in line with the S.U.N ratio 5.1\% of the base model on space group No.225. This nevertheless demonstrates that the reinforcement fine-tuning procedure can guide the generative process toward discovering candidates within the pretrained dataset with no property label.
Notably, two materials in Table~\ref{table:dielectric_samples} are \emph{absent} from the Alex-20 dataset, highlighting the ability of RL to guide the generative process toward novel samples beyond the pretrained dataset.

\begin{table*}[htbp]
%\resizebox{\columnwidth}{!}{
\begin{threeparttable}
\centering
\caption{Dielectric materials discovered by \texttt{CrystalFormer-RL}, ranked according to their figures of merit (FoM): $\varepsilon_{\mathrm{elec}} \cdot E_{g}$. The band gap, dielectric constant, and energy above the convex hull are all calculated by DFT. The two highlighted rows are not in the Alex-20 dataset.
See details of the samples at \href{https://zenodo.org/records/17982949}{https://zenodo.org/records/17982949}.
}
\renewcommand{\arraystretch}{1.2}
\setlength{\tabcolsep}{10pt}
\begin{tabular}{@{}lrrrr@{}}
\toprule
Formula & $E_g$ (eV)~\tnote{1} &  $\varepsilon_{elec}$ & FoM (eV)& $E_{\mathrm{hull}}$ (eV/atom)  \\
\midrule
\rowcolor{yellow} Cs$_2$LiLuF$_6$ & 7.37 & 2.38 & 17.52 & 0.09  \\
Cs$_2$LiTmF$_6$ & 7.35 & 2.37 & 17.40 & 0.08   \\
Cs$_2$LiHoF$_6$ & 7.32 & 2.37 & 17.39 & 0.09   \\
Cs$_2$LiErF$_6$ & 7.29 & 2.37 & 17.28 & 0.09   \\
Cs$_2$LiTbF$_6$ & 7.21 & 2.36 & 17.05 & 0.09   \\
Cs$_2$LiDyF$_6$ & 7.17 & 2.34 & 16.80 & 0.09   \\
CsCaF$_3$ & 6.94 & 2.32 & 16.10 & 0.06   \\
Cs$_2$LiScF$_6$ & 6.64 & 2.40 & 15.92 & 0.09   \\
Cs$_2$NaTmF$_6$ & 6.98 & 2.27 & 15.84 & 0.05   \\
Cs$_2$NaErF$_6$ & 6.96 & 2.27 & 15.77 & 0.05   \\
Cs$_2$NaLuF$_6$ & 6.93 & 2.26 & 15.63 & 0.05   \\
Cs$_2$NaHoF$_6$ & 6.90 & 2.26 & 15.59 & 0.05   \\
Cs$_2$NaYF$_6$ & 6.89 & 2.25 & 15.48 & 0.03  \\
Cs$_2$LiLuCl$_6$ & 5.34 & 2.90 & 15.48 & 0.01  \\
\rowcolor{yellow}
Rb$_2$LiLuF$_6$ & 7.01 & 2.19 & 15.34 & 0.07   \\
Cs$_2$LiHoCl$_6$ & 5.16 & 2.86 & 14.78 & 0.01   \\
Cs$_2$LiTbCl$_6$ & 5.07 & 2.89 & 14.66 & 0.01  \\
Y$_2$ErF$_6$ & 7.15 & 1.96 & 14.01 & 0.05   \\
Cs$_2$SrMgCl$_6$ & 5.07 & 2.74 & 13.89 & 0.03   \\
Cs$_2$MgPbF$_6$ & 5.29 & 2.60 & 13.77 & 0.08   \\
K$_2$Rb$_2$AlF$_6$ & 6.40 & 2.02 & 12.92 & 0.06  \\
Cs$_2$RbDyCl$_6$ & 4.92 & 2.46 & 12.09 & 0.03   \\
Cs$_2$ErGaCl$_6$ & 3.82 & 3.05 & 11.64 & 0.04   \\
Cs$_2$BaPbI$_6$ & 3.17 & 3.62 & 11.48 & 0.07  \\
Cs$_2$NaAlCl$_6$ & 3.97 & 2.77 & 11.00 & 0.03  \\
Rb$_2$TmGaCl$_6$ & 3.60 & 2.99 & 10.77 & 0.06  \\
\bottomrule
\label{table:dielectric_samples}
\end{tabular}
% \begin{tablenotes}
%     \item[1] The band gap and total dielectric constant are predicted by machine learning models.
%     \item[2] The energy above the convex hull is computed via the Orb model based on Alexandria convex hull.
% \end{tablenotes}
\end{threeparttable}
\end{table*}

\begin{figure}[htbp]
    \centering
    \includegraphics[width=1\linewidth]{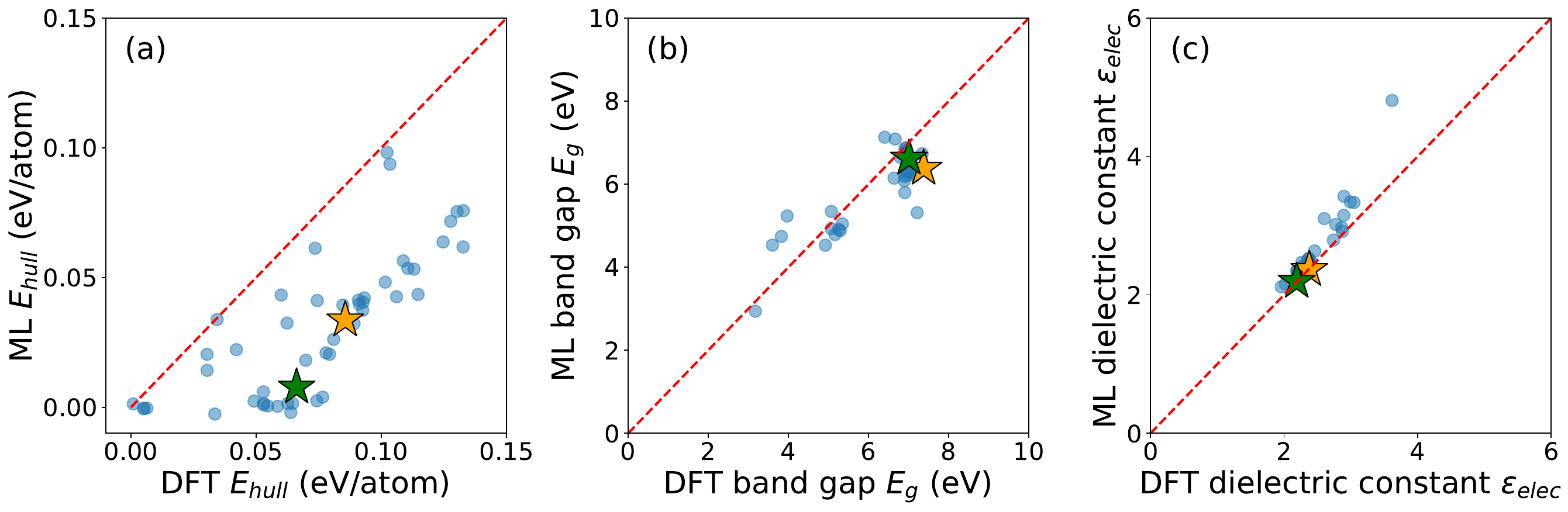} 
    \caption{
        ML predicted vs. DFT calculated (a) energy above the convex hull $E_{\mathrm{hull}}$, (b) band gap $E_g$ and (c) electronic dielectric constant $\varepsilon_{\mathrm{elec}}$
        for the discovered dielectric materials. Panels (b) and (c) include only materials with $E_{\mathrm{hull}} < 0.1$ eV/atom according to the DFT calculation. The two stars indicate the same two novel samples as Fig.~\ref{fig:dielectric}(a).
    }
    \label{fig:dft}
\end{figure}

We compare the DFT-calculated band gap and electronic dielectric constant of the generated samples with the predictions from the MEGNet and AnisoNet models. As shown in Figure~\ref{fig:dft}, the ML predictions align well with the DFT results, validating the effectiveness of the RL fine-tuning approach for property-guided material generation.

% We compare the prediction results of the generated samples with that of corresponding equilibrium structures from the Alex-20 dataset in Table~\ref{table:alex_dielectric_samples}. We observe that the generated materials are close to the equilibrium structure in the Alex-20 dataset, with similar predicted FoM values.

The \texttt{CrystalFormer}'s ability to aid property-guided exploration of crystalline materials has been demonstrated in previous work~\cite{cao2024space} using the MCMC sampling method from the posterior distribution according to the Bayes rule. 
Both the MCMC and reinforcement fine-tuning approach work with any machine learning interatomic potential or property prediction models trained separately of the \texttt{CrystalFormer}. 
The RL approach addresses the long-mixing time or even the ergodicity issue of the MCMC sampling of the posterior distribution with the additional effort of fine-tuning and a variationally approximated posterior. 

\section{Discussions}
\label{sec:discussion}

Reinforcement fine-tuning of materials generative model is useful for optimizing other materials properties such as the thermoelectric figure of merit, thermal conductivity, and superconducting critical temperature. In addition to a foundational generative model such as the \texttt{CrystalFormer} for the materials prior $p(x)$, the framework requires a reliable ML prediction model for the likelihood $p(y|x)$, which is a research topic with rapid progress both on generalizability and accuracy, see, e.g.~\cite{pota2024thermalconductivitypredictionsfoundation, loew2024universalmachinelearninginteratomic,fu2025learning}. 

It is often easier to discriminate the utility of a given material than to generate a material with the desired utility. Reinforcement fine-tuning of a materials generative model using discriminative feedback offers a way to bridge such discrimination-generation gap. RL on the autoregressive transformer such as \texttt{CrystalFormer}~\cite{cao2024space} is particularly convenient given a variety of post-training techniques available for LLM. Although those RL techniques are also applicable to natural language based materials generative models~\cite{flam2023language, antunes2024crystal, gruver2024finetuned}, we believe finetuned \texttt{CrystalFormer} is more advantageous because it natively speaks the language of crystalline materials, which cherishes the fundamental inductive bias of crystalline materials such as space group symmetry and periodicity. Furthermore, the idea of reinforcement finetuning can also be applied to the material generative models based on diffusion~\cite{xu2025design, chen2025matinvent}.
The diffusion model is typically fine-tuned through the incorporation of auxiliary networks to condition the generation on external inputs~\cite{zhang2023adding}, while RL eliminates the necessity for these additional network components.
For example, Refs.~\cite{fan2024optimizing, black2024training} employ the RL algorithm to fine-tune the pre-trained text-to-image diffusion model, leading to improved sample quality and diversity. In the mean time, applying RL with KL regularization to the diffusion model is challenging due to the need for precise log-likelihood calculations~\cite{fan2023dpok}.

It is also interesting to compare active learning and RL in the context of generative materials discovery. Active learning~\cite{xiao2025invdesflow, li2025activelearningconditionalinverse} selects crystal samples back to the generative training dataset, which is similar to the SFT approach discussed in Sec.~\ref{sec:stability}. Similar to pretrain, the active learning procedure optimizes the forward KL divergence between the empirical distribution of the training dataset and the generative model. In contrast, RL minimizes the reverse KL divergence~\cite{korbak2022rl} between the generative model and the target distribution \Eq{eq:optimal_policy}. There are both empirical and theoretical evidences that the RL approach delivers better performance compared to the active learning approach in the contexts of LLM fine-tuning~\cite{shao2024deepseekmath, shenfeld2025rlsrazoronlinereinforcement}.
Consequently, the primary advantage of RL over active learning is its ability to directly incorporate task-specific reward signals,
enabling the model to generate outputs that more effectively meet practical application requirements. Moreover, we suspect the difference in the mode covering versus mode seeking in the active learning versus RL may also be a reason for the difference in the performance~\cite{Goodfellow-et-al-2016, levine2018reinforcement}.

Materials generative models are typically trained on equilibrium or near equilibrium structures, which only represent a small fraction of the chemical space. Incorporating non-equilibrium structures into the training process can improve model robustness and enhance the model's capability to generate novel structures. Reinforcement fine-tuning achieves this by using MLIP reward model to penalize the unstable structures generated by the model. %Exploring the design of the reward function can lead to different outcomes, which presents an avenue for further investigation. \lw{what does this sentence intend to say ? Looks to me redundant.}
Alternatively, the direct preference optimization~\cite{rafailov2024direct} algorithm may be employed to fine-tune the model on pairs of equilibrium and non-equilibrium structures, which eliminates the necessity of training an explicit reward model for material stability. 

% \lw{this paragraph should be repharazed since we already performed DFT calculation for the Ehull} \zdcao{Fig 3 and Fig 6b both use Orb model to evaluate the Ehull. We did not use DFT to evaluate the Ehull for all generated samples.}
% The stability verifications presented in Fig.~\ref{fig:ehull} and Fig.~\ref{fig:dielectric}(b) are based on the energy above the convex hull evaluated with an MLIP. 
Figure~\ref{fig:dft} compares the ML-predicted and DFT-calculated results, revealing that while the overall trends are consistent, noticeable discrepancies remain between the predictions and the DFT evaluations.
This reveals a fundamental challenge in the application of RL in materials science: what is the verifiable ground truth reward signal. Given the intricacies involved in material synthesis and characterization~\cite{cheetham2024artificial}, such a challenge is particularly concerning as faulty reward functions~\cite{clark2016faulty} which fails to align with the final goal will suggest structures with high reward but fail for the final experiment or even intermediate DFT verifications.  
We believe the design of reward functions that are both actionable and well-aligned with experimental outcomes, with the insights of human experts, is crucial for ensuring the reliability of the present and many related efforts. 

Compared to property-guided material generation based on training on labeled data~\cite{gomez2018automatic, zeni2023mattergen, ye2024cdvae, luo2024deep}, the approaches advertised in~\cite{cao2024space} (MCMC) and the present work (Reinforcement fine-tuning, also known as variational Bayes) rely on the external discriminative model in a plug-and-play manner. The flexibility and modularity nature of this approach allows seamless integration with many existing and very strong MLIP and ML property prediction models into the materials generative design workflow. 

This study focuses on how generative models can be enhanced and specialized through the integration of property prediction models.
However, the reverse question: how property prediction models can be improved by leveraging generative models—is also an intriguing question worth future investigation. For example, the Orb models employ a two-stage training framework that integrates generative and supervised learning to achieve state-of-the-art performance~\cite{neumann2024orb}. Initially, the model is pre-trained as denoising diffusion models on datasets of stable materials. The pre-trained model then serves as the initialization for an MLIP, which is subsequently fine-tuned in a supervised manner to predict the energy, forces, and stress of materials. This approach exemplifies how generative pre-training can enhance the performance of discriminative models such as MLIP. Extending this idea further, Ref.~\cite{zhang2023dpa2} has demonstrated that fine-tuning pretrained MLIP also gives better performance for property prediction models. 

%\lw{maybe Orb is already an example ? First train the score function of diffusion model, than tune it to a force field model. }

%\lw{discuss the danger of reward hacking, namely some of the samples that are good for orb may note be stable in the DFT sense. (pushing that to a limit, even DFT Ehull is good does not mean that it is good). material science is in the end an open domain.}

\begin{acknowledgments}
We thank Han Wang, Linfeng Zhang, Jian Lv, Hongteng Xu, Shigang Ou, Xiaoshan Luo, Mingzhang Yang, Tianping Ying, Ling Lu, Weiluo Ren and Huaxian Jia for useful discussions. This project is supported by the National Natural Science Foundation of China under Grants No. T2225018, No. 92270107, No. 12188101, and No. T2121001, Cross-Disciplinary Key Project of Beijing Natural Science Foundation No. Z250005, National Key Projects for Research and Development of China Grant. No. 2021YFA1400400, and the Strategic Priority Research Program of Chinese Academy of Sciences under Grants No. XDB0500000. 
\end{acknowledgments}

\bibliography{refs}
\clearpage

\onecolumngrid

\setcounter{algorithm}{0}
\renewcommand{\thealgorithm}{\arabic{algorithm}}
\setcounter{table}{0}
\renewcommand{\thetable}{S\arabic{table}}
\setcounter{figure}{0}
\renewcommand{\thefigure}{S\arabic{figure}}
\setcounter{equation}{0}
\renewcommand{\theequation}{S\arabic{equation}}

\appendix

\section{The \texttt{CrystalFormer} model card}

The architecture of \texttt{CrystalFormer} is identical to the one described in Ref.~\cite{cao2024space}, except for an update on the positional encoding. We use the rotary positional encoding (RoPE)~\cite{su2023roformer} instead of the absolute positional encoding in the original design.
% In fact, crystal data exhibits a multimodal nature, suggesting the potential utilization of multimodal 2D-RoPE~\cite{heo2024rotary} for 
% enhanced position encoding, which could capture the diverse structural characteristics. \lw{then why not using 2D-RoPE now? To avoid question you many omit this sentence.}

The base model was trained for 5500 epochs with a total batch size of 8000 over eight A100 GPUs using the Adam optimizer~\cite{kingma2017adam}. The learning rate was set to $8e-3$.

The reinforcement fine-tuning was performed for 100 - 200 steps, depending on the task. For the reinforcement fine-tuning, we use a total batch size of 1000 over two A100 GPUs and the Adam optimizer with a learning rate of $1e-4$. The pre-training phase required approximately 1000 GPU hours, while the reinforcement fine-tuning phase took around 12 to 24 GPU hours.

% Gradient clipping was applied by value at 1 for all the stages of training. \lw{move information of this sentence into the table, something like grad\_clip 1}

For the supervised fine-tuning (SFT) of the \texttt{CrystalFormer}, we generate 10,000 crystal samples and relax them using the Orb models. Only those stable samples with $E_{\mathrm{hull}}<0.1$ eV/atom were retained for SFT. The SFT was performed for 55 epochs when validation loss stopped improving, 
with a batch size of 100 using the Adam optimizer with a learning rate of $1e-5$ on a single A100 GPU. 

\begin{table*}[h!]
\caption{A table of hyperparameters used in this work.}
\begin{center}
\renewcommand{\arraystretch}{1.2}
\setlength{\tabcolsep}{24pt} 
\begin{tabular}{lcl}
\toprule
\textbf{Hyperparameters} & \textbf{Value} & \textbf{Remarks} \\
\hline
The length of atom sequence including the padding atoms & 21 \\
Number of chemical species  &  119  & 'H' to 'Og', plus padding atom \\
Number of possible Wyckoff letters & 28 & 'a-z'+'A', plus padding atom  \\
Number of modes in von-Mises mixture distribution $K_{x}$ & 16 \\
Number of modes in lattice Gaussian mixture distribution $K_{l}$ & 4  \\
Hidden layer dimension for the Wyckoff letter of the first atom & 256 \\
Transformer number of layers & 16\\
Transformer number of heads & 16 \\
Transformer key size & 64 \\
Transformer model size  $d_\mathrm{model}$ & 64 \\
Embedding dimension of discrete input  & 32 \\
Number of Fourier frequency $N_{f}$ & 5 \\
Dropout rate in pre-training & 0.1 & No dropout in fine-tuning \\ 
Gradient clipping & 1 \\
\hline
RL regularization coefficient $\tau$ & 0.1 & 2 for FoM reward \\
PPO clipping parameter $\epsilon$ & 0.2 \\
PPO steps between resampling & 3 \\
\hline
Total number of parameters:  4833575 \\
\bottomrule
\end{tabular}
\end{center}
\label{table:hyperparameters}
\end{table*}

% \begin{table*}[h!]
% \caption{A table of hyperparameters used in RL.}
% \begin{center}
% \renewcommand{\arraystretch}{1.2}
% \setlength{\tabcolsep}{24pt} 
% \begin{tabular}{lc}
% \toprule
% \textbf{Hyperparameters} & \textbf{Value} \\
% \hline
% regularization coefficient $\tau$ & 0.1 \\
% PPO clipping parameter $\epsilon$ & 0.2 \\
% PPO steps between resampling & 3 \\
% \bottomrule
% \end{tabular}
% \end{center}
% \label{table:hyperparameters-rl}
% \end{table*}

\newpage
\section{Training dataset}
\label{sec:dataset-app}

The Alexandria dataset~\cite{schmidt2024improving} was chosen as a starting point because it was the largest openly available DFT dataset of equilibrium 
and near equilibrium structures ($\sim$ 4.5 million materials). Using the same criteria as MatterGen~\cite{zeni2023mattergen} (distances to the convex hull less than 0.1 eV/atom and less than 20 atoms in the unit cell), Alexandria 
now contains more than 1.3 million materials. 80 \% of the data set was used for training and the remaining were divided equally
between validation and test datasets. The Alex-20 dataset is available at~\cite{alex20}.

\begin{figure}[htbp]
    \centering
    \includegraphics[width=1\linewidth]{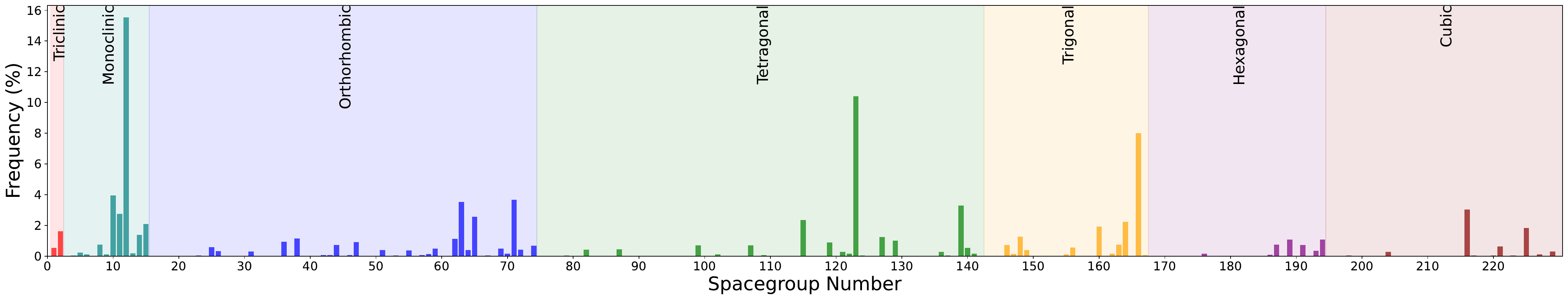} 
    \caption{Space group distribution of the Alex-20 dataset.}
    \label{fig:spg_distribution}
\end{figure}

\begin{figure}[htbp]
    \centering
    \includegraphics[width=0.8\linewidth]{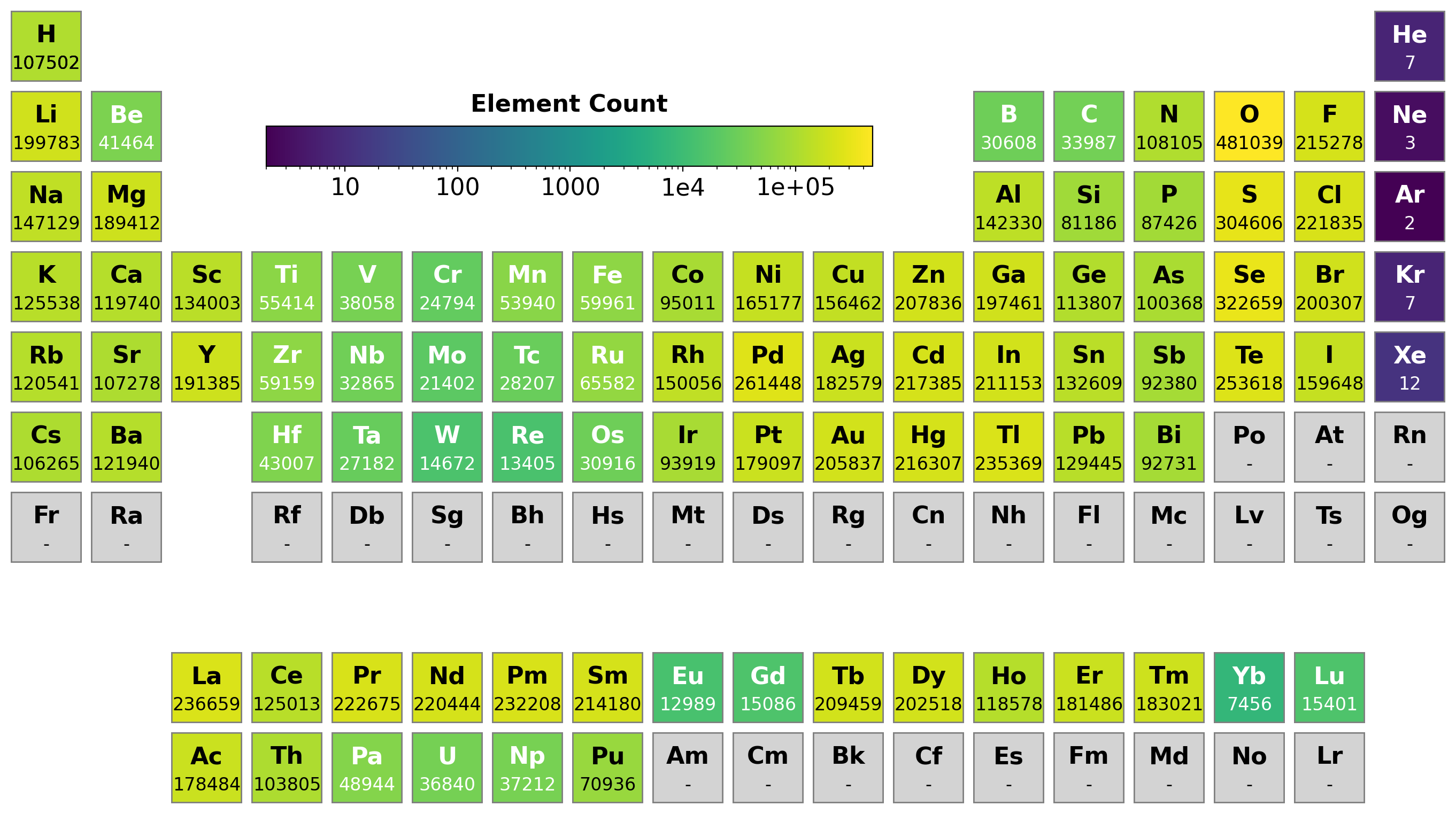} 
    \caption{Distribution of elements in Alex-20.}
    \label{fig:element_distribution}
\end{figure}

\begin{figure}[htbp]
    \centering
    \includegraphics[width=0.5\linewidth]{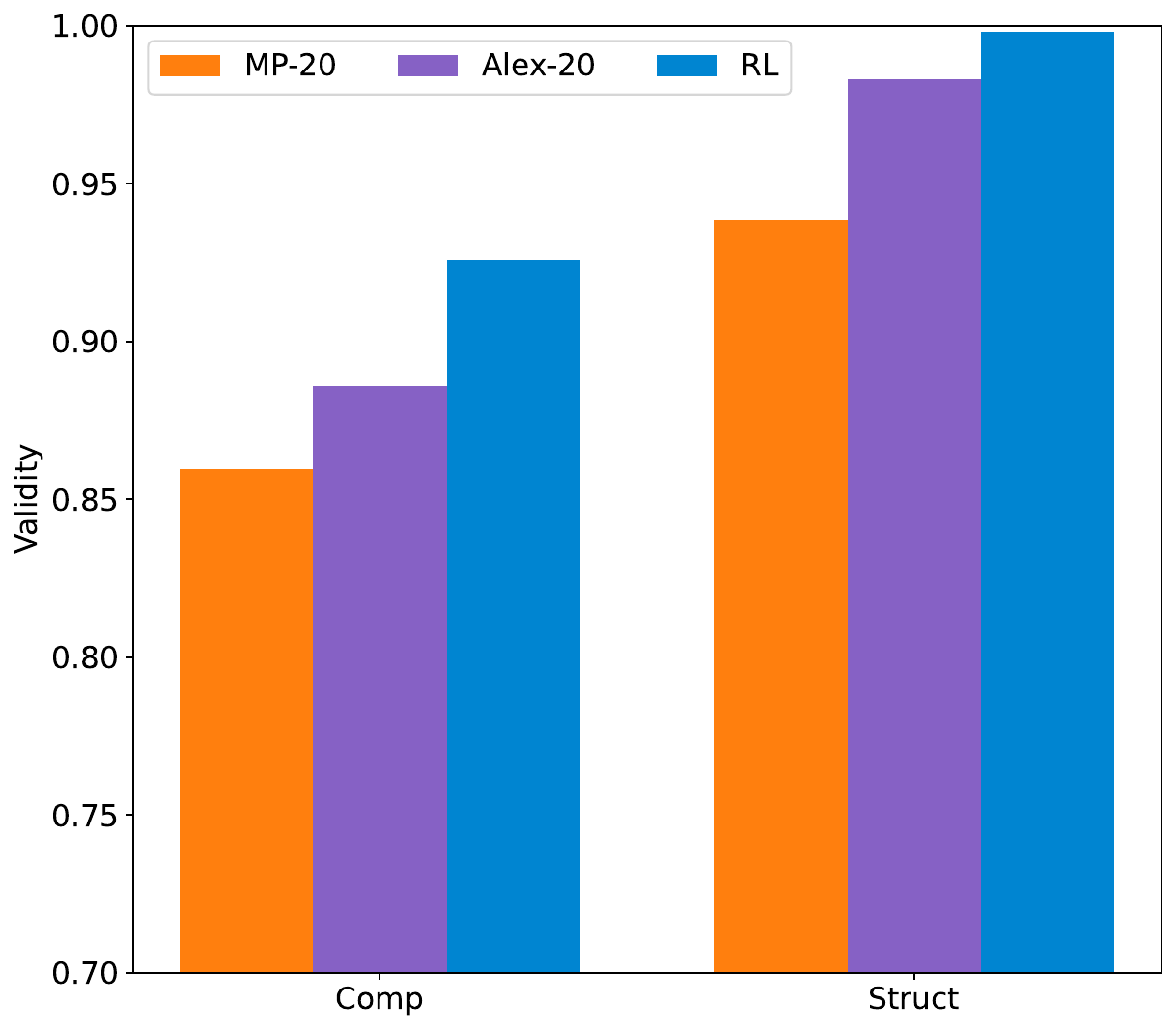} 
    \caption{Composition and structure validity of samples generated by the pretrained \texttt{CrystalFormer} using the MP-20 and Alex-20 dataset, as well as reinforcement fine-tuned model using the energy above hull as the reward function. 
    }
    \label{fig:validity}
\end{figure}
\clearpage

\newpage
\section{Details of reinforcement fine-tuning implementation}
\label{sec:ppo-app}
The proximal policy optimization (PPO)~\cite{schulman2017proximal} is a simple yet effective policy gradient-based RL algorithm. The algorithm optimizes the following objective function,
\begin{equation}
    \mathcal{L}^{\mathrm{PPO}}(\theta) = \mathop{\mathbb{E}}_{x \sim p_{\mathrm{old}}(x)} \left [  \mathrm{min}(\gamma(\theta)A(x), \mathrm{clip}(\gamma(\theta), 1-\epsilon, 1+\epsilon)A(x) ) \right ],
\end{equation}
where $\gamma(\theta)=\frac{p_{\theta}(x)}{p_{\mathrm{old}}(x)}$ denotes the probability ratio between the policy under optimization and the policy for sampling, $A(x)$ is the advantage function, and $\epsilon$ is the clipping parameter.
The second term, $\mathrm{clip}(\gamma(\theta), 1-\epsilon, 1+\epsilon)$, modifies the surrogate objective by clipping the probability ratio, which removes the incentive for moving $\gamma(\theta)$ outside of the interval $[1-\epsilon, 1+\epsilon]$.
Drawing samples from $p_{\mathrm{old}}(x)$ rather than $p_{\theta}(x)$ enables sample reuse and allows multiple gradient update steps on the same batch of data, thereby improving sample efficiency without requiring costly re-evaluation of the reward function for each update.

The advantage function $\mathcal{A}(x)$ is then defined as
\begin{equation}
    \mathcal{A}(x) = r(x) - b,
\end{equation}
where $b$ represents the baseline. For simplicity, we incorporate the KL regularization term into the advantage function, yielding
\begin{equation}
    A(x) = \mathcal{A}(x) - \tau \ln \frac{p_{\theta}(x)}{p_{\mathrm{base}}(x)}.
\end{equation}
Note that one does not need to compute the gradient of $A(x)$ with respect to network parameters $\theta$ in the PPO algorithm. Instead, one uses the score function gradient estimate~\cite{mohamed2020monte} to maximize the objective function $\mathcal{L}^{\mathrm{PPO}}(\theta)$. For each batch of samples from $p_{\mathrm{old}}$ and reward signal, we carry out a few gradient ascent steps to update the policy network. The detailed hyperparameters used in the PPO algorithm are listed in Table~\ref{table:hyperparameters}. 

The advantage function measures the deviation of the reward $r(x)$ from this baseline, indicating how much better or worse the generated samples is compared to the baseline.
In standard PPO implementations, this baseline is typically approximated by a value function, which is trained alongside the policy model. We use the average reward of the generated samples $x$, i.e., $b=\mathop{\mathbb{E}}_{x\sim p_{\mathrm{old}}}[r(x)]$ as the baseline~\cite{williams1992simple}.
This approach obviates the need for training a separate value function and is closely related to the group relative policy optimization done right~\cite{shao2024deepseekmath, liu2025understanding}. 

For the baseline $b$ we use an exponential moving average of the previous rewards, which stabilizes the training and also leverages the past reward information to form a lagged baseline.
To be more specific, the baseline is updated as $b_{k} = \eta b_{k-1} + (1-\eta)\bar{R}_{k}$, where $\eta=0.95$ and $b_{0}$ is initialized as the average reward of the initial batch $\bar{R}_{0}$. Here, $\bar{R}_{k}$ is the average reward of the current batch.

\newpage
\section{Computational details}
\label{sec:computational-details}

\subsection{Machine learning force field}
We employ the Orb-v2~\cite{neumann2024orb} trained on the 
MPTraj~\cite{deng2023chgnet} and Alexandria dataset~\cite{schmidt2024improving} to as the reward model for energy prediction for the generated samples. 
We calculate the energy above the convex hull using the energy predicted by the Orb model in the Materials Project energy correction framework (i.e., \texttt{MaterialsProject2020Compatibility} from pymatgen~\cite{ong2013python}).

\subsection{DFT calculations}

The DFT calculations were performed with the Perdew-Burke-Ernzerhof (PBE) exchange-correlation functional~\cite{Perdew.PBE.1996} and all-electron projector-augmented wave method~\cite{Blochl.PAW.1994}, as implemented in the VASP code~\cite{Kresse.VASP.1996}.
All parameters of the calculations including settings of PBE functional, Hubbard U corrections, and ferromagnetic initialization are chosen to be consistent with Materials Project by using of \texttt{\detokenize{MPRelaxSet}} function in pymatgen~\cite{ong2013python}.
All structures containing Yb element are ignored when calculating energy above the hull due to they are unavailable from the Materials Project at the time of writing.
The DFPT input files are generated using the \texttt{\detokenize{MPStaticSet}} in pymatgen, with a k-point density of 3,000 per reciprocal atom and a plane-wave energy cutoff of 600 eV, consistent with the Materials Project settings~\cite{petousis2017high}.

\end{document}